\documentclass{amsart}

\usepackage{epsfig}
\usepackage{graphicx}
\usepackage{amsmath,amssymb,amsfonts,latexsym}
\usepackage{graphicx}

\usepackage{amssymb}
\usepackage{enumerate, xspace}
\usepackage{dsfont}
\usepackage{afterpage}
\usepackage{changebar}
\usepackage{amsthm}
\usepackage{rotating}

\numberwithin{equation}{section}

\begin{document}

\title[Quantum adiabatic theorem for chemical reactions]
  {Quantum adiabatic theorem for chemical reactions and systems with time-\\dependent orthogonalization}
\author[Andrew Das Arulsamy]{Andrew Das Arulsamy} \email{sadwerdna@gmail.com}

\address{Condensed Matter Group, Division of Interdisciplinary Science, D403 Puteri Court, No.1, Jalan 28, Taman Putra, 
68000 Ampang, Selangor DE, Malaysia}




\date{\today}

\begin{abstract} 
A general quantum adiabatic theorem with and without the time-dependent orthogonalization is proven, which can be applied to understand the origin of activation energies in chemical reactions. Further proofs are also developed for the oscillating Schwinger Hamiltonian to establish the relationship between the internal (due to time-dependent eigenfunctions) and external (due to time-dependent Hamiltonian) timescales. We prove that this relationship needs to be taken as an independent condition for the quantum adiabatic approximation. We give four examples, including logical expositions based on the spin-$\frac{1}{2}$ two-level system to address the gapped and gapless (due to energy level crossings) systems, as well as to understand how does this theorem allows one to study dynamical systems such as chemical reactions. 
\end{abstract}

\maketitle



\section{Introduction} 

The quantum adiabatic theorem (QAT) and its approximation (QAA) have been the backbone in many areas of quantum physics, namely, in condensed matter theory (via the Born-Oppenheimer approximation)~\cite{oppen}, in atoms, molecules and quantum chemistry~\cite{lind}, in quantum field theory via the Gell-Mann and Low theorem~\cite{gell}, and presently in the adiabatic quantum computation~\cite{farhi}. The theorem was first discussed by Ehrenfest~\cite{ehrenfest}, and later was formally derived by Born and Fock~\cite{bornf} and Kato~\cite{kato}. Other modern proofs for QAT can be found in Refs.~\cite{ste,griffiths5,messiah,comp,avron,berry,sab}. We refer the readers to Refs.~\cite{ste,comp,avron} for thorough reviews on the development of QAT. Apart from the historical perspectives, Comparat~\cite{comp} have also proven the existence of two sufficient QAT conditions for the oscillating Schwinger Hamiltonian. Teufel-Spohn~\cite{ste2,ste3} and Avron-Elgart~\cite{avron} on the other hand, have provided the proof for the validity of QAT for systems obeying the Born-Oppenheimer approximation, and for gapless systems, respectively. We will revisit the two conditions given by Comparat to prove that these conditions are in fact equivalent to the relationship between $T_{\rm ex}$ and $T_{\rm in}$. Here, $T_{\rm ex}$ and $T_{\rm in}$ are the respective external and internal timescales. Subsequently, we will discuss how the new theorem proven here addresses the gapless system. Due to numerous applications of QAT in physics, a proper theorem with explicitly defined conditions is needed. Here, we prove the existence of such a theorem, which can also be exploited to understand the origin of chemical reactions. To arrive at the standard QAT for gapped system, we first let $H(t)$, $\psi(t)$ and $E(t)$ to be the time-dependent Hamiltonian, eigenfunction and eigenvalue, respectively, such that~\cite{griffiths5} 
\begin {eqnarray}
H(t)\psi_n(t) = E_n(t)\psi_n(t) \label{eq:new1},
\end {eqnarray}
and $\{\psi_n(t)\}$ constitutes a complete orthonormal set that satisfies
\begin {eqnarray}
\langle\psi_n(t)|\psi_m(t)\rangle = \delta_{nm} \label{eq:new2},
\end {eqnarray}
in which $\psi_n(t)$ can be expressed as a linear combination 
\begin {eqnarray}
\Psi(t) = \sum_n c_n(t)\psi_n(t)e^{i\theta_n(t)} \label{eq:77.2},
\end {eqnarray}
with the phase factor 
\begin {eqnarray}
\theta_n(t) = -\frac{1}{\hbar}\int_0^tE_n(t')dt', \label{eq:new4}
\end {eqnarray}
and $\{n,m\} \in \mathbb{N^*}$. We use $\mathbb{N}^*$ and $\mathbb{R}$ to denote the set of positive integers excluding zero and the set of real numbers, respectively. Here, Eq.~(\ref{eq:77.2}) is the general solution to the time-dependent Schr$\ddot{\rm o}$dinger equation
\begin {eqnarray}
i\hbar\frac{\partial}{\partial t}\Psi(t) = H(t)\Psi(t). \label{eq:77.3}
\end {eqnarray}
Therefore, one can show that if the initial Hamiltonian, $H_{\rm init}(t)$ evolves very slowly to a new final Hamiltonian, $H_{\rm fin}(t)$, then a particle that starts out in the $n^{\rm th}$ eigenstate ($\Psi^{\rm init}_n(t) = \psi_n(t)e^{i\theta_n(t)}$) of $H_{\rm init}(t)$ is expected to occupy the ground state ($\Psi_n^{\rm fin}(t) = \psi_n(t)e^{i\gamma_n(t)}e^{i\theta_n(t)}$) of $H_{\rm fin}(t)$. Observe that the final eigenstate has non-trivially, picked up only a harmless phase factor
\begin {eqnarray}
\gamma_n(t) = i\int_0^t\bigg\langle\varphi_n(t')\bigg|\frac{\partial}{\partial t'}\varphi_n(t')\bigg\rangle dt'. \label{eq:neww4}
\end {eqnarray}
In the above formulation, Griffith used Eqs.~(\ref{eq:77.2}) and~(\ref{eq:77.3}) to obtain the required coefficient (in explicit form)~\cite{griffiths5},
\begin {eqnarray}
&&\dot{c}_m(t) = -c_m(t)\langle \varphi_m(t)|\dot{\varphi}_m(t)\rangle - \sum_{n\neq m} c_n(t) \frac{\left\langle \varphi_m(t)\left|\dot{H}(t)\right|\varphi_n(t)\right\rangle}{E_n(t) - E_m(t)} e^{i(\theta_n (t) - \theta_m (t))}, \nonumber \\&& \label{eq:77.5}
\end {eqnarray}
that can be used to find out how slow the evolution should be, so that the initial particle still occupies the ground state of the new (final) Hamiltonian. Here, $c_m(t)$ and $c_n(t)$ denote the $m^{\rm th}$ and $n^{\rm th}$ eigenstate coefficients, respectively, such that $|c_m(t)|^2 + |c_n(t)|^2 = 1$. From Eq.~(\ref{eq:77.5}), we can notice that both $E_n(t)$ and $E_m(t)$ exist at all times, and their eigenfunctions are orthogonal. This means that if $E_n(t)$ and $E_m(t)$ are degenerate at certain point of time, then Eq.~(\ref{eq:77.5}) cannot be used to derive the adiabatic criterion. The reason is that the gap that exists when $t = t_0$ does not exist for $t = t_1$. Now, assuming non-degeneracy, the QAT criterion is given by  
\begin {eqnarray}
\left|\frac{\left\langle \varphi_m(t)\left|\dot{H}(t)\right|\varphi_n(t)\right\rangle}{E_n(t) - E_m(t)}\right| \ll \dot{c}_m(t), \label{eq:77.1}
\end {eqnarray}
which needs to be satisfied for one to exploit QAA. Here, $\varphi_m(t)$ is the usual $t$-dependent eigenfunction for the $m^{\rm th}$ eigenstate and $E_n(t)$ is the $t$-dependent eigenvalue for the $n^{\rm th}$ eigenstate, and the dot represents time derivative. There are two important assumptions associated to the original QAT given in Eq.~(\ref{eq:77.1}). First, there is no quantum phase transitions or chemical reactions when the initial quantum system evolves from $t_0$ to $t_1$. Secondly, the QAT is only applicable for a two- or a multi-level system with well-defined energy gaps (strictly no energy level crossings). For example, when one considers a two-level system for simplicity, then there exist two $t$-dependent eigenfunctions, observable at all times, and they are always nondegenerate and the two eigenvalues are $E_n(t)$ $\neq$ $E_m(t)$. In fact, we can add as many eigenfunctions as we possibly could (as denoted by $n$ and $m$), but we chose two, to keep the discussion straightforward. Of course, later we will go through the examples with large number of energy levels and energy level crossings. 

For the sake of argument however, if we assume that $E_n(t = t_0)$ $\neq$ $E_n(t = t_1)$ (this is definitely different from $E_n(t = t)$ $\neq$ $E_m(t = t)$, as described earlier) then we cannot apply the quantum adiabatic approximation [given in Eq.~(\ref{eq:77.1})] simply because both of these eigenvalues ($E_n(t = t_0)$ and $E_n(t = t_1)$) are not observable simultaneously. Recall that $t_0 < t_1$. Furthermore, $E_n(t = t_0)$ $\neq$ $E_n(t = t_1)$ implies the energy gap ($g$) in the form of $g = |E_n(t = t_0) - E_n(t = t_1)|$. The energy gap in this form \textit{cannot} be defined as an energy gap because 
\begin {eqnarray}
&&\omega'(t) = \frac{E_n(t = t_0) - E_n(t = t_1)}{\hbar} \neq \nonumber \\&& \omega (t)_{\rm{Bohr}} = \frac{E_n(t) - E_m(t)}{\hbar} = \frac{1}{T_{\rm in}(t)}. \label{eq:77.7}
\end {eqnarray}    
In other words, physically valid energy gaps must satisfy $\omega (t)_{\rm{Bohr}}$. Later, we will show that $\omega'(t)$ can be associated to the possibility of a particle traveling forward and backward in time. Now, it is also important to note here that if the external disturbances with a timescale, 
\begin {eqnarray}
\frac{\hbar\omega_{\rm Bohr}}{T_{\rm ex}(t)} \propto \left\langle \varphi_m(t)\left|\dot{H}(t)\right|\varphi_n(t)\right\rangle \label{eq:77.7tex}, 
\end {eqnarray}
approaches $\omega_{\rm Bohr}$ due to $T_{\rm ex} \approx T_{\rm in}$, then one cannot apply QAA. Now, one should not be carried away into thinking that Eq.~(\ref{eq:77.1}) lacks a small parameter, which is required to understand the quantum adiabatic evolution. For example, the smallness is usually controlled by the dilation operator, $\tau$ such that $s_{\rm ex} = T_{\rm ex}/\tau$. In our formalism however, this smallness is controlled by the relationship between $T_{\rm in}$ and $T_{\rm ex}$, without the need to introduce any \textit{ad-hoc} control parameter \textit{by hand}. We will get to this point in detail in Example 1 when we proof Eq.~(\ref{eq:77.7tex}). The adiabatic criteria, $T_{\rm ex} \gg T_{\rm in}$ and Eq.~(\ref{eq:77.1}) have been proven in Ref.~\cite{griffiths5}, however, without the proper definitions for these timescales. The existence of $\omega'(t)$ means that we are equating $\varphi_n(t)$ with $\varphi_n(t = t_0)$ and $\varphi_m(t)$ with $\varphi_n(t = t_1)$ in Eq.~(\ref{eq:77.1}). In this case, we are referring to only one eigenfunction at all times such that $\varphi_n(t = t_0)$ evolves to $\varphi_n(t = t_1)$, where $\varphi_n(t = t_0)$ and $\varphi_n(t = t_1)$ may or may not be orthogonal to each other. If they are orthogonal, then one obtains $\omega'(t)$. On the other hand, if they are not orthogonal, then $\omega'(t)$ does not exist. 

In addition to our arguments based on Eq.~(\ref{eq:77.7}), Comparat~\cite{comp} have also shown that the condition given in Eq.~(\ref{eq:77.1}) is valid provided that the Hamiltonian is real, gapped and non-oscillating. However, the generalized conditions derived there~\cite{comp}, which are valid for any Hamiltonian, do not consider gapless systems nor the Hamiltonians that allow $t$-dependent orthogonalization. In addition, the internal and external timescales are also not properly taken into account. The $t$-dependent orthogonalization and large changes to the Hamiltonians (due to large external disturbances) are essential because they can lead to continuous or second-order quantum phase transitions (QPT) during chemical reactions. The readers are referred to Refs.~\cite{sach,sach2,sen,sen2} for further details on the concept of QPT and its applications in strongly correlated matter. 

We anticipate that this continuous QPT proposed by Sachdev~\cite{sach,sach2} and Senthil~\cite{sen,sen2} can also occur due to the existence of energy level crossings due to large external disturbances, with or without the $t$-dependent orthogonalization, which could be the origin of any chemical reaction between two chemical species. This means that the energy levels of unreacted species must cross by overcoming the activation energies to form new compounds, and these new compounds (if formed) will have a new set of energy levels or bands. If the formed compounds are of non free-electronic systems, then there is a theorem based on the ionization energy theory that states$-$ the energy level spacing of such a system is proportional to their constituent atomic energy level spacing~\cite{phd}. Physically, this means that the energy level spacing of this newly formed compound is proportional to the energy level spacing of their constituent chemical elements. 

Before we move on, let us first try to understand the concept of internal and external timescales in the QAT. It is well known now that there are different versions of QATs available for different applications as reviewed by Teufel~\cite{ste}, and Avron-Elgart~\cite{avron}. However, all the existing QATs are solely based on the scaled time, $s_{\rm ex} = t_{\rm ex}/\tau_{\rm ex}$, $s_{\rm ex} \in [0,1]$, without explicitly referring to the internal and external timescales. In particular, $s_{\rm ex}$ here is independent of the internal timescale. The reason for this scenario is that there are no available equations to properly define $T_{\rm ex}$ and $T_{\rm in}$ independently. Anyway, the scaled (dimensionless) time, $s_{\rm ex}$ contains the dilation time or a time dilation operator, $\tau_{\rm ex}$, such that $\tau_{\rm ex}$ allows us to slow down the time $t_{\rm ex}$ to obtain a slowly varying Hamiltonian with respect to the value 1 (when $\tau_{\rm ex} = t_{\rm ex}$). Hence, $t_{\rm ex}$ is the original (unscaled) physical time for the Hamiltonian, and $\tau_{\rm ex}$ determines the dilation of this external physical time. 

In the unscaled formalism, we simply write $T_{\rm ex}$ to denote external timescale, to avoid confusion with $t$-dependence in $H(t)$ and $\varphi(t)$ (eigenfunction). Otherwise, we will have to write $H(T_{\rm ex})$ and $\varphi(T_{\rm in})$, and this notation will confuse the readers into thinking that both $H(T_{\rm ex})$ and $\varphi(T_{\rm in})$ are not normalized to the real-space physical time, $t$, or $\varphi(T_{\rm in})$ is from a different Hamiltonian, and so on. As a consequence, we can see that the Hamiltonian with scaled time, $H(s_{\rm ex})$ does not take the internal timescale into account, which can only be taken into account by defining the time variation in the wave functions or the eigenfunctions for a given $H(s_{\rm ex})$. Similar to $s_{\rm ex}$, we can also define $s_{\rm in} = t_{\rm in}/\tau_{\rm in}$, where $s_{\rm in} \in [0,1]$. In these scaled notations, $t_{\rm in}$ refers to the original physical time of the wave functions, and $\tau_{\rm in}$ determines the dilation of the internal physical time. Here, $s_{\rm in}$ is independent of the external timescale. Again, in the unscaled formalism, we write $T_{\rm in}$ to denote internal timescale. 

Having understood the different notations used in the scaled and unscaled formalisms, we can now properly define the time, $T_{\rm ex}$ as the external timescale (characteristic time for changes in the Hamiltonian), while $T_{\rm in}$ is the internal timescale (characteristic time for changes in the wave function or eigenfunction). In order to understand the existence of these properly defined timescales, we do the following: we can define $t_{\rm in} = t_{\rm ex} = t$ and write $s_{\rm in} = t/\tau_{\rm in} = t/T_{\rm in}$ and $s_{\rm ex} = t/\tau_{\rm ex} = t/T_{\rm ex}$. This normalization transforms $H(s_{\rm ex})$ back to $H(t)$, and the only thing left to do here is to find the inequality relation between $T_{\rm ex}$ and $T_{\rm in}$. However, this $H(s_{\rm ex})$ or $H(t)$ still ignores the internal timescale (because it is fixed as a constant), which leads to an important question$-$ how small should $\tau_{\rm ex}$ (or $T_{\rm ex}$) be with respect to $T_{\rm in}$ in order to violate the QAA? This question means that we cannot claim a given Hamiltonian varies slowly if $\tau_{\rm ex} \rightarrow \infty$ because we have to make sure $\tau_{\rm in}$ does not go to $\infty$ when $\tau_{\rm ex} \rightarrow \infty$. We will prove here that the relation between $T_{\rm in}$ and $T_{\rm ex}$ needs to be treated as a separate condition altogether.

Now we will discuss why the concept of quantum adiabaticity is correct, and why any violation that may exist only indicate the non-applicability of the QAA for certain systems. The existence of non-orthogonality in strongly interacting physical systems (or non-linear quantum systems) has been discussed by Yukalov~\cite{yuka} as a result of non-linear terms in a given Hamiltonian that destroys the adiabatic evolution. These non-linear terms can originate from various interactions such as electron-electron, electron-phonon, spin-orbit, spin-spin and so on, which may give rise to numerous energy-level crossings along the time line, $t$. As anticipated, this means that we need to keep track of the orthogonality of the eigenfunctions with respect to time (which will be explained in the subsequent sub-section). It is to be noted here that strongly correlated systems, including quantum dots and nanostructures (non-free-electron systems) do not form gapless systems. For example, the electrons and the phonons cannot be decoupled, unless $T_{\rm electron} \ll T_{\rm phonon}$~\cite{phys}. 

Any violation to a given QAT means the existence of a non-adiabatic process, and this does not imply the QAT is incorrect. Violating the QAT deliberately is important as a crosscheck because if the QAT is violated for adiabatic processes, then the QAT is either (a) not applicable for that process (for example, applying gapless or gapped QAT to a system that has time($t$)-dependent orthogonalization), or (b) the said QAT is insufficient (again, this does not mean the concept of quantum adiabaticity is incorrect). Consequently, deliberately violating a given QAT can be used as a mean to crosscheck the correctness and sufficiency of the conditions given in a particular QAT, as well as to have a deeper understanding of the conditions needed to guarantee the QAA, for example, with respect to the existence of self-orthogonality~\cite{lefeb}.  

Here, we develop a new version of the QAT that (i) allows $t$-dependent orthogonalization for each single eigenfunction as an additional condition, which can be used to explain the possibility of a particle traveling backward in time, (ii) allows to understand the origin of chemical reactions in the absence of, and in the presence of $t$-dependent orthogonalization, and (iii) gives a proper definition for the $T_{\rm ex}$ and $T_{\rm in}$ and establishes the relationship between them as a valid and separate condition. Points (i) to (iii) are the main results of this paper. In addition, we apply our theorem to degenerate energy levels due to energy level crossings to validate the QAA with properly defined $T_{\rm ex}$ and $T_{\rm in}$ without invoking any Rabi oscillation.

\section{Generalized Quantum Adiabatic Theorem}

We use $\mathcal{H}$ for the complex Hilbert space, while $||\varphi||$ is the norm of an eigenfunction ($\varphi$). \\

\textbf{Theorem 1.} \textit{The quantum eigenfunctions} (\textit{or wave functions}) \textit{that satisfy the quantum mechanical postulates}~\cite{andrei} \textit{are represented by the orthonormalized complex vectors}, $|\varphi(t)\rangle = (\cdots,|\varphi(t_j)\rangle,\cdots,|\varphi(t_k)\rangle,\cdots,|\varphi(t_z)\rangle)$, \textit{where} $|\varphi(t)\rangle \in \mathcal{H}$ and $||\varphi(t)||^2 = \langle \varphi(t)|\varphi(t)\rangle = \int_{\mathbb{R}^{z}}|\varphi(t)|^2dt = 1$. \textit{If a quantum system with a time-dependent Hamiltonian, $H_{\rm init}(t)$ is initially $(t = t_j)$ in its ground state $(|\varphi_n(t)\rangle$ or $|\varphi_m(t_j)\rangle)$, evolves into a final $(t = t_k)$ quantum system with a Hamiltonian, $H_{\rm fin}(t)$ then the probability for the final quantum system to be in its ground state $(|\varphi_m(t)\rangle$ or $|\varphi_m(t_k)\rangle)$ is controlled by these three criteria}
\begin {eqnarray}
&&\left|\frac{\left\langle \varphi_{m}(t_k)\left|\dot{H}\right|\varphi_{m}(t_j)\right\rangle - \dot{E}_{m}(t_j)\langle \varphi_{m}(t_k)|\varphi_{m}(t_j)\rangle}{E_{m}(t_j) - E_{m}(t_k)}\right| \approx \dot{c}_m(t_k), \label{eq:77.1new}
\end {eqnarray}
\begin {eqnarray}
\left|\frac{\left\langle \varphi_{m}(t)\left|\dot{H}(t)\right|\varphi_{n}(t)\right\rangle}{E_{n}(t) - E_{m}(t)}\right| < \dot{c}_m(t), ~~~\rm {and}\label{eq:77.1new1} 
\end {eqnarray}
\begin {eqnarray}
T_{\rm ex}(t) > T_{\rm in}(t). \label{eq:x2}
\end {eqnarray}
\textit{Here} $t \in [t_j,t_k]$ $\in$ $\mathbb{R}$, \textit{and} $\left\langle \varphi_{m}(t_k)|\varphi_{m}(t_j)\right\rangle = \delta_{t_kt_j} = 0$ \textit{implies} $\varphi_{m}(t_j)$ \textit{is orthogonal to} $\varphi_{m}(t_k)$. \textit{On the other hand}, $\left\langle \varphi_{m}(t_k)|\varphi_{m}(t_j)\right\rangle = \delta_{t_kt_j} = 1$ \textit{implies} $\varphi_{m}(t_j)$ \textit{is not orthogonal to} $\varphi_{m}(t_k)$ and $\{j,k,m,n\} \in \mathbb{N}^*$. 

\textit{Remark 1}: Equation~(\ref{eq:77.1new}) keeps track of all the individual $t$-dependent eigenfunctions in a given system. QAA is strictly valid if (\textbf{a}): $\approx \dot{c}_m(t_k)$ in Eq.~(\ref{eq:77.1new}) is replaced with $= \dot{c}_m(t_k)$, (\textbf{b}): $< \dot{c}_m(t)$ in Eq.~(\ref{eq:77.1new1}) is replaced with $\ll \dot{c}_m(t)$, and (\textbf{c}): $>$ in Eq.~(\ref{eq:x2}) is replaced with $\gg$. In fact, we can reversibly switch to strict QAA conditions if needed, or loosen them, or use them to invalidate the QAA in certain systems. We can now write a statement in regards to QAT: The probability for a particle that was initially in the ground state of an initial Hamiltonian, $H_{\rm init}(t)$, to occupy any excited state of some final Hamiltonian, $H_{\rm fin}(t)$ is approximately zero, if $H_{\rm init}(t)$ changes slowly to its final form, $H_{\rm fin}(t)$ such that Eqs.~(\ref{eq:77.1new}),~(\ref{eq:77.1new1}) and~(\ref{eq:x2}) are satisfied. 

\textit{Remark 2}: Time discretizations have been carried out to obtain instantaneous eigenfunctions. In other words, the $t$-dependent eigenfunction has been time-discretized in the form of $\varphi_m(t) = \cdots, \varphi_{m}(t_j), \cdots, \varphi_{m}(t_k)\cdots, \varphi_{m}(t_z)$. Time discretizations are carried out in pairs ($t_j$ and $t_k$) to obtain two time-discretized eigenfunctions ($\varphi_{m}(t_j)$ and $\varphi_{m}(t_k)$) so as to check their orthogonality. The sole purpose of this time-discretization is to capture the $t$-dependent orthogonalization and/or the wave function transformation (that will be exposed in the last section). 

\textit{Remark 3}: In addition, it is to be noted here that $c(t_j) = c(t = t_j)$, $\varphi(t_j) = \varphi(t = t_j)$ and $\theta(t_j) = \theta(t = t_j)$. As for the time derivative variables, $\dot{c}(t_j) = \frac{\partial c(t)}{\partial t}(t = t_j)$, $\dot{\varphi}(t_j) = \frac{\partial\varphi(t)}{\partial t}(t = t_j)$, and so on.    

\textbf{Proof} for Eq.~(\ref{eq:77.1new}): The validity of Eqs.~(\ref{eq:77.1new1}) and~(\ref{eq:x2}) have been proven in Ref.~\cite{griffiths5}. The discussion on how Eq.~(\ref{eq:77.1new}) is related to chemical reactions is given in Example 4, while the proof for Eq.~(\ref{eq:77.7tex}) is given in Example 1. We will start with a single eigenfunction ($m = 1$) for two discretized times, i.e., when $t = t_j$ and $t = t_k$. For simplicity, we let $j = 0$ and $k = 1$ from here onwards, and then we can just repeat this procedure for $j = 1$ and $k = 2$, $j = 2$ and $k = 3$, and so on. Moreover, $\varphi_{m}(t_0)$ is allowed to evolve and to be orthogonal or not orthogonal to $\varphi_{m}(t_1)$ when $t = t_1$. The $t$-dependent Hamiltonian, say for $t = t_0$ is given by 
\begin {eqnarray}
H(t_0)\varphi(t_0) = E(t_0)\varphi(t_0). \label{eq:77.10}
\end {eqnarray}
The $t$-dependent Schr$\ddot{\rm o}$dinger equation is given in Eq.~(\ref{eq:77.2}) in which, $\Psi(t)$ can be expressed as a linear combination of $\varphi(t_0)$ and $\varphi(t_1)$ (from Eq.~(\ref{eq:77.3})) 
\begin {eqnarray}
\Psi(t) = \sum^{t_1}_{t_0} c(t_0)\varphi(t_0) e^{i\theta(t_0)} ~\Leftrightarrow ~t\in [t_0,t_1]. \label{eq:77.9}
\end {eqnarray}
The summation is used between $t_0$ and $t_1$ because we have discretized the $t$-dependent eigenfunction (See Remark 2). Here, we also have $c(t_0)$, which is a coefficient that is related to the probability of finding a particle at a given point of time ($t_0$ or $t_1$), and with a particular eigenvalue, $E_{m}(t_0)$ or $E_{m}(t_1)$, respectively, for $m$ = 1 or 2 or $\cdots$. For example, Eq.~(\ref{eq:77.1new}) implies $|c_{m}(t_0)|^2 + |c_{m}(t_1)|^2 = 1$, while Eq.~(\ref{eq:77.1new1}) implies $|c_{m}(t)|^2 + |c_{n}(t)|^2 = 1$ (for $m \neq n$). Recall here that we are only considering a single eigenfunction ($m$ = 1), and due to Remark 2, it is convenient for us to drop the subscript $m$. From Eqs.~(\ref{eq:77.2}) and~(\ref{eq:77.9})
\begin {eqnarray}
&&\dot{\Psi} = \sum^{t_1}_{t_0} \big[\dot{c}(t_0)\varphi(t_0) + c(t_0)\dot{\varphi}(t_0) + ic(t_0)\varphi(t_0)\dot{\theta}(t_0) \big]e^{i\theta(t_0)}, \nonumber \\&&
i\hbar\sum^{t_1}_{t_0} \big[\dot{c}(t_0)\varphi(t_0) + c(t_0)\dot{\varphi}(t_0) + ic(t_0)\varphi(t_0)\dot{\theta}(t_0) \big]e^{i\theta(t_0)} \nonumber \\&& = \sum^{t_1}_{t_0}c(t_0)H\varphi(t_0)e^{i\theta(t_0)}. \label{eq:78AA}
\end {eqnarray}
Using  
\begin {eqnarray}
\theta(t_0) = -\frac{1}{\hbar}\bigg[\int_0^{t}E(t')dt'\bigg]_{t = t_0} ~\Rightarrow ~ \dot{\theta}(t_0) = -\frac{1}{\hbar}E(t_0), \label{eq:79AA}
\end {eqnarray}
we obtain
\begin {eqnarray}
&&i\hbar\sum^{t_1}_{t_0} \big[\dot{c}(t_0)\varphi(t_0) + c(t_0)\dot{\varphi}(t_0)\big]e^{i\theta(t_0)} \nonumber \\&& + i\hbar\sum^{t_1}_{t_0}ic(t_0)\varphi(t_0)\bigg(-\frac{1}{\hbar}E(t_0)\bigg)e^{i\theta(t_0)} = \sum^{t_1}_{t_0}c(t_0)E(t_0)\varphi(t_0)e^{i\theta(t_0)}, \nonumber \\&& i\hbar\sum^{t_1}_{t_0} \big[\dot{c}(t_0)\varphi(t_0) + c(t_0)\dot{\varphi}(t_0)\big]e^{i\theta(t_0)} \nonumber \\&& + \sum^{t_1}_{t_0}c(t_0)E(t_0)\varphi(t_0)e^{i\theta(t_0)} = \sum^{t_1}_{t_0}c(t_0)E(t_0)\varphi(t_0)e^{i\theta(t_0)}, \label{eq:80AA}
\end {eqnarray}
which leads to 
\begin {eqnarray}
\sum^{t_1}_{t_0} \dot{c}(t_0)\varphi(t_0) e^{i\theta(t_0)} = -\sum^{t_1}_{t_0} c(t_0)\dot{\varphi}(t_0) e^{i\theta(t_0)}. \label{eq:77.11}
\end {eqnarray}    
Now comes the crucial part, we will take the inner product with $\varphi(t_1)$, which is an eigenfunction at a later time, $t_1$ ($t_1 > t_0$), evolved from $\varphi(t_0)$. Therefore,
\begin {eqnarray}
&&\langle\varphi(t_1)|\sum^{t_1}_{t_0} \dot{c}(t_0)|\varphi(t_0)\rangle e^{i\theta(t_0)} = -\langle\varphi(t_1)|\sum^{t_1}_{t_0} c(t_0)|\dot{\varphi}(t_0)\rangle e^{i\theta(t_0)}, \label{eq:81AA}
\end {eqnarray}  
where
\begin {eqnarray}
\left\langle \varphi(t_1)|\varphi(t_0)\right\rangle = \delta_{t_1t_0}. \label{eq:77.12}
\end {eqnarray}    
From Eq.~(\ref{eq:77.12}), one obtains 
\begin {eqnarray}
\sum^{t_1}_{t_0} \dot{c}(t_0) \delta_{t_1t_0} e^{i\theta(t_0)} = -\sum^{t_1}_{t_0} c(t_0)\left\langle \varphi(t_1)|\dot{\varphi}(t_0)\right\rangle e^{i\theta(t_0)}. \label{eq:77.13}
\end {eqnarray}    
Now, let us assume that the evolved eigenfunction, $\varphi(t_1)$ is \textit{not} orthogonal to $\varphi(t_0)$ or $\delta_{t_1t_0} = 1$. Hence, Eq.~(\ref{eq:77.13}) can be written as  
\begin {eqnarray}
&&\dot{c}(t_1)\delta_{t_1t_0}e^{i\theta(t_1)} = -\sum^{t_1}_{t_0} c(t_0)\left\langle \varphi(t_1)|\dot{\varphi}(t_0)\right\rangle e^{i\theta(t_0)} \nonumber \\&& \dot{c}(t_1) = -\sum^{t_1}_{t_0} c(t_0)\left\langle \varphi(t_1)|\dot{\varphi}(t_0)\right\rangle e^{i[\theta(t_0)-\theta(t_1)]}. \label{eq:77.14}
\end {eqnarray}    
Differentiating Eq.~(\ref{eq:77.10}), taking the inner product with $\varphi(t_1)$ and using $\langle\varphi(t_1)|H|\dot{\varphi}(t_0)\rangle = E(t_1)\langle\varphi(t_1)|\dot{\varphi}(t_0)\rangle$ we can derive 
\begin {eqnarray}
&&\frac{\partial}{\partial t}(H\varphi(t_0)) = \frac{\partial}{\partial t}(E(t_0)\varphi(t_0)), \nonumber \\&& \dot{H}\varphi(t_0) + H\dot{\varphi}(t_0) = \dot{E}(t_0)\varphi(t_0) + E(t_0)\dot{\varphi}(t_0), \nonumber \\&&  \langle \varphi(t_1)|\dot{H}|\varphi(t_0)\rangle + E(t_1)\langle\varphi(t_1)|\dot{\varphi}(t_0)\rangle = \dot{E}(t_0) + E(t_0)\langle \varphi(t_1)|\dot{\varphi}(t_0)\rangle. \label{eq:82AA}
\end {eqnarray}    
Since the single eigenfunction at $t_0$ remains as a single eigenfunction at $t_1$, there is no available excited eigenfunctions for any transition to occur. Therefore the transition probability is simply zero. Invoking $\delta_{t_1t_0} = 1$, we can rewrite Eq.~(\ref{eq:82AA}) as
\begin {eqnarray}
(E(t_0) - E(t_1))\langle \varphi(t_1)|\dot{\varphi}(t_0)\rangle = \langle \varphi(t_1)|\dot{H}|\varphi(t_0)\rangle - \dot{E}(t_0). \label{eq:77.15}
\end {eqnarray}    
Substituting Eq.~(\ref{eq:77.15}) into Eq.~(\ref{eq:77.14}), we obtain 
\begin {eqnarray}
&&\dot{c}(t_1) = -c(t_1)\langle \varphi(t_1)|\dot{\varphi}(t_1)\rangle -\sum_{t_0\neq t_1} c(t_0)\left\langle \varphi(t_1)|\dot{\varphi}(t_0)\right\rangle e^{i[\theta(t_0)-\theta(t_1)]} \nonumber \\&& = -c(t_1)\left\langle \varphi(t_1)|\dot{\varphi}(t_1)\right\rangle - \sum_{t_0 \neq t_1} c(t_0)\frac{\left\langle \varphi(t_1)\left|\dot{H}\right|\varphi(t_0)\right\rangle - \dot{E}(t_0)}{E(t_0) - E(t_1)} e^{i[\theta(t_0)-\theta(t_1)]}. \label{eq:77.16}
\end {eqnarray}    
Equation~(\ref{eq:77.16}) is straightforward where $c(t_0) = c(t_1) = 1$ because there is only one eigenfunction, and is not orthogonal along the time line for $t \in [t_0,t_1]$. Let us now invoke the orthogonality ($\delta_{t_1t_0} = 0$), which implies that (from Eq.~(\ref{eq:77.14}))
\begin {eqnarray}
&& 0 = -\sum^{t_1}_{t_0} c(t_0)\left\langle \varphi(t_1)|\dot{\varphi}(t_0)\right\rangle e^{i\theta(t_0)}. \label{eq:77.14a}
\end {eqnarray}    
Using Eqs.~(\ref{eq:77.14a}) and~(\ref{eq:82AA}), 
\begin {eqnarray}
&&c(t_1) = -\frac{c(t_0)e^{i\theta(t_0)}}{\langle \varphi(t_1)|\dot{\varphi}(t_1)\rangle}\sum_{t_0 \neq t_1} \frac{\left\langle \varphi(t_1)\left|\dot{H}\right|\varphi(t_0)\right\rangle}{E(t_0) - E(t_1)}, \label{eq:77.16a}
\end {eqnarray}    
which in turn allows us to conclude that 
\begin {eqnarray}
\left|\frac{\left\langle \varphi(t_1)\left|\dot{H}\right|\varphi(t_0)\right\rangle}{E(t_0) - E(t_1)}\right| \approx 1 \Rightarrow c(t_1) \approx 1, ~~~\rm {or}\label{eq:77.17}
\end {eqnarray}
\begin {eqnarray}
\left|\frac{\left\langle \varphi(t_1)\left|\dot{H}\right|\varphi(t_0)\right\rangle}{E(t_0) - E(t_1)}\right| \ll 1 \Rightarrow c(t_1) \ll 1. \label{eq:77.17new}
\end {eqnarray}
Equations~(\ref{eq:77.17}) and~(\ref{eq:77.17new}) are equal to the adiabatic criterion given in Eq.~(\ref{eq:77.1new}) if $\dot{E}(t_0)\langle \varphi(t_1)|\varphi(t_0)\rangle \neq 0$ $\blacksquare$ 

As indicated earlier, systems with a single eigenfunction, and in the presence of $t$-dependent orthogonalization, we will face the consequence of a particle traveling backward in time due to $c(t_1) \ll 1$ (Eq.~(\ref{eq:77.17new})) in accordance with $\omega'(t)$ (see Eq.~(\ref{eq:77.7})). This is physically not possible because we will not be able to orthogonalize (assuming this can be done) the eigenfunction without disturbing the occupied electron. In other words, it is not possible because the eigenfunction here itself represents the wave function of the electron based on the Hermitian quantum mechanics~\cite{ord}. Therefore, Eq.~(\ref{eq:77.17new}) strictly implies that the ground state electron has been excited or ionized. We cannot be sure how this (excitation or ionization) is possible from the criterion given in Eq.~(\ref{eq:77.1new}) alone, it could be due to degeneracy or invalid Eq.~(\ref{eq:x2}) [for example $T_{\rm ex}(t) < T_{\rm in}(t)$]. Later in Example 4 we will justify that this phenomenon ($c(t_1) \ll 1$) can be related to chemical reactions (with and without the $t$-dependent orthogonalization).    

\section{Applications of Theorem 1}

Now we are ready to discuss the applications of Theorem 1, and how this theorem relates to these examples both mathematically and physically. The first example is for spin-$1/2$ two-level system, which is by far the most straightforward one. The second one deals with the same spin-$1/2$ two-level system but with more and complicated $t$-dependent external disturbances. Subsequently, we discuss gapless system due to energy level crossings, while the last one invokes the $t$-dependent orthogonalization in the spin-$1/2$ two-level system via Eq.~(\ref{eq:77.1new}) to check the applicability of QAA for such systems. We also show that Eq.~(\ref{eq:77.1new}) can be exploited to understand chemical reactions in the absence of $t$-dependent orthogonalization.    

\subsection*{Example 1: Spin-$\frac{1}{2}$ two-level system in the presence of magnetic field}

\begin{figure}
\begin{center}
\scalebox{0.2}{\includegraphics{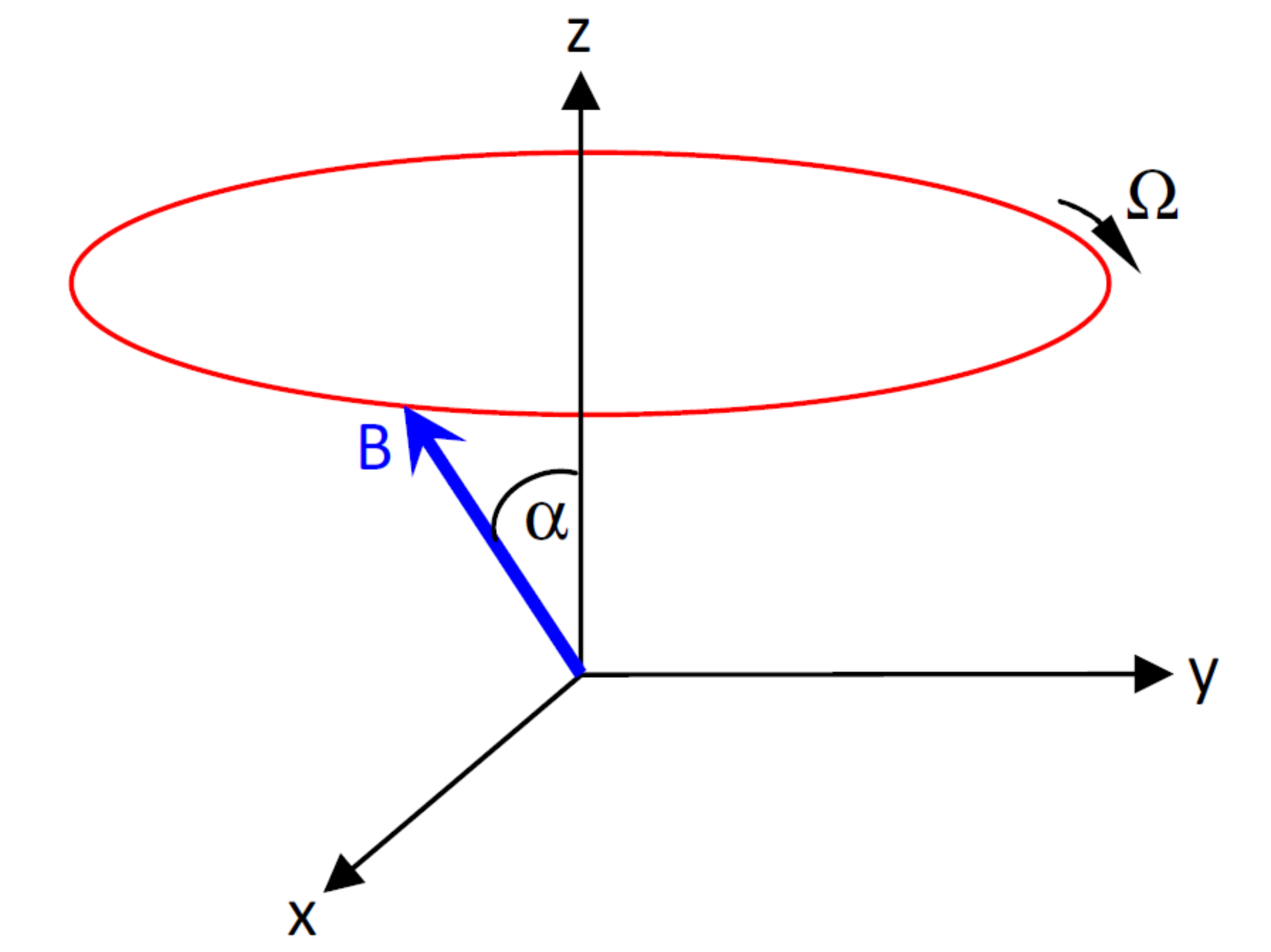}}
\caption{A constant magnetic field, $\textbf{B}$ (magnitude $B_0$) rotates at a constant velocity, $\Omega$, and with a constant opening angle, $\alpha$. Here, $\textbf{B}(t) = B_0[\sin\alpha\cos(\Omega t)\hat{\textbf{i}} + \sin\alpha\sin(\Omega t)\hat{\textbf{j}} + \cos\alpha \hat{\textbf{k}}]$. See Example 1 for details.}
\label{fig:1}
\end{center}
\end{figure}

The first example is the one derived by Griffith~\cite{griffiths5}. An electron can have spin up or down, and therefore, in the presence of rotating (constant velocity = $\Omega$) and constant magnetic field with a constant opening angle, $\alpha$, one can control the transition probability of this electron (from spin up to down) to satisfy~\cite{griffiths5}
\begin {eqnarray}
|\langle\Psi(t)|\varphi_{\rm down}(t)\rangle|^2 = \bigg[\frac{\Omega}{\Lambda}\sin\alpha\sin\bigg(\frac{\Lambda t}{2}\bigg)\bigg]^2, \label{eq:xx1}
\end {eqnarray}
where $\Lambda^2 = \Omega^2 + \omega^2_{\rm Bohr} - 2\Omega\omega_{\rm Bohr}\cos\alpha$ and $\omega_{\rm Bohr} = [E_{\rm up} - E_{\rm down}]/\hbar = 1/T_{\rm in}$. Obviously, in this example we have $\Omega = 1/T_{\rm ex}$. In the limit $T_{\rm ex} \gg T_{\rm in}$ due to $\Omega \ll \omega_{\rm Bohr}$, one obtains the transition probability to spin down to be zero because $\Lambda \rightarrow \omega_{\rm Bohr}$. Therefore, ($\Omega/\Lambda$) $\rightarrow$ ($\Omega/\omega_{\rm Bohr}$) $\rightarrow$ 0 in accordance with Eq.~(\ref{eq:xx1}), satisfying Eq.~(\ref{eq:x2}) in Theorem 1. In this physical example, we do not need Eq.~(\ref{eq:77.1new}) as it should be, because it is a gapped system without any $t$-dependent orthogonalization.   
In order to prove Eq.~(\ref{eq:77.7tex}), we write the Hamiltonian~\cite{griffiths5} for this system (see Fig.~\ref{fig:1})

\[ H_{\rm G}(t)
= \frac{\hbar\omega_{\rm Bohr}}{2}\begin{pmatrix}
\cos\alpha & e^{-i\Omega t}\sin\alpha  \\
e^{i\Omega t}\sin\alpha & -\cos\alpha   
\end{pmatrix},\] 
and its normalized eigenspinors~\cite{griffiths5} 
\[ \varphi_{\rm up}(t)
= \begin{pmatrix}
\cos(\alpha/2) \\
e^{i\Omega t}\sin(\alpha/2)
\end{pmatrix}~~~\rm and~~~ \varphi_{\rm{down}}(\textsl{t})
= \begin{pmatrix}
e^{-i\Omega t}\sin(\alpha/2)\\
-\cos(\alpha/2) 
\end{pmatrix}.\]
Their corresponding eigenvalues are $\hbar\omega_{\rm Bohr}/2$ and $-\hbar\omega_{\rm Bohr}/2$, respectively. From Eq.~(\ref{eq:77.2}), the exact solution~\cite{griffiths5} 
\[
\Psi(t) 
= \begin{pmatrix} 
\bigg[\cos(\Lambda t/2) - \big[i(\omega_{\rm Bohr} - \Omega)/\Lambda\big] \sin(\Lambda t/2) \bigg]\cos(\alpha/2)e^{-i\Omega t/2} \\
\\ \bigg[\cos(\Lambda t/2) - \big[i(\omega_{\rm Bohr} + \Omega)/\Lambda\big] \sin(\Lambda t/2) \bigg]\sin(\alpha/2)e^{i\Omega t/2}
\end{pmatrix}.\]
\textbf{Proof} for Eq.~(\ref{eq:77.7tex}): We now evaluate 
\begin {eqnarray}
\left\langle \varphi_{\rm down}(t)\left|\dot{H}(t)\right|\varphi_{\rm up}(t)\right\rangle &=& \Omega\bigg[\frac{i}{2}\sin\alpha \bigg]\hbar\omega_{\rm Bohr}e^{-i\Omega t}. \label{eq:xxx0}
\end {eqnarray}
Since $\omega_{\rm Bohr}$ is a constant and $\Omega = 1/T_{\rm ex}$, we can readily obtain the proportionality given in Eq.~(\ref{eq:77.7tex}) $\blacksquare$ \\
If $\omega_{\rm Bohr}$ is $t$-dependent, then one obtains
\begin {eqnarray} 
\frac{\left\langle \varphi_{\rm down}(t)\left|\dot{H}(t)\right|\varphi_{\rm up}(t)\right\rangle}{E_{\rm up}(t) - E_{\rm down}(t)} \propto \frac{1}{T_{\rm ex}(t)} \ll 1. \label{eq:xxx00}
\end {eqnarray}
Equation~(\ref{eq:xxx00}) implies that $T_{\rm ex} \gg 1$ for QAA to be valid, and this inequality does not refer to $T_{\rm in}$ explicitly as explained in the introduction. It is not explicit because we can rewrite Eq.~(\ref{eq:xxx00}) to obtain  
\begin {eqnarray} 
\left\langle \varphi_{\rm down}(t)\left|\dot{H}(t)\right|\varphi_{\rm up}(t)\right\rangle \propto \frac{\hbar/T_{\rm in}(t)}{T_{\rm ex}(t)} \propto \frac{\hbar}{T_{\rm in}(t) T_{\rm ex}(t)} \ll 1, \label{eq:xxx00z}
\end {eqnarray}
which means that we are still not able to find the relationship between $T_{\rm in}(t)$ and $T_{\rm ex}(t)$. Therefore, Eq.~(\ref{eq:x2}) needs to be invoked independent of Eq.~(\ref{eq:77.1new1}).

\subsection*{Example 2: Spin-$\frac{1}{2}$ two-level oscillating system}

\begin{figure}
\begin{center}
\scalebox{0.2}{\includegraphics{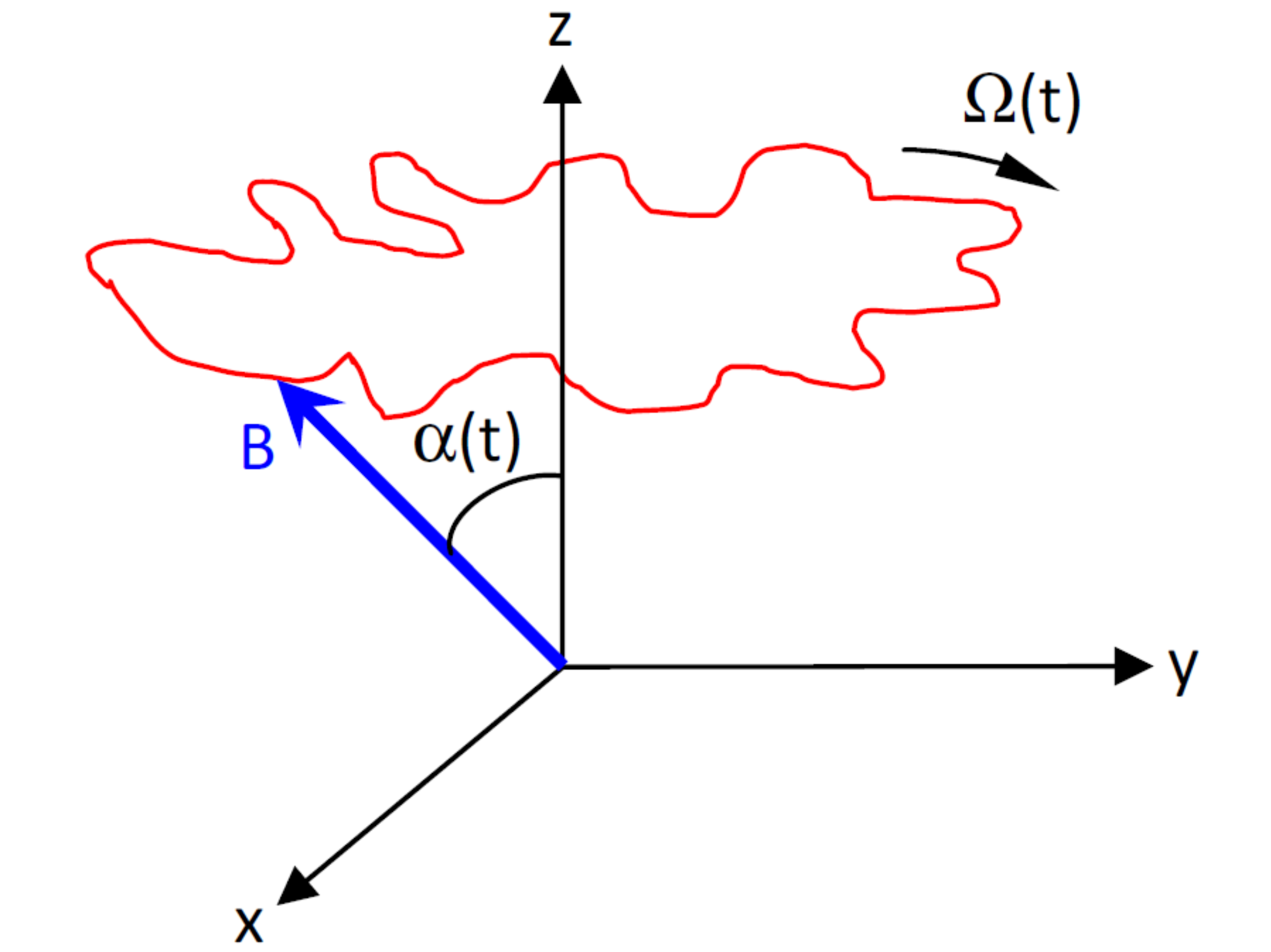}}
\caption{A constant magnetic field, $\textbf{B}$ (magnitude $B_0$) rotates with some fluctuations at a velocity, $\Omega(t)$ and with a $t$-dependent opening angle, $\alpha(t)$. See Example 2 for details.}
\label{fig:2}
\end{center}
\end{figure}

In the above example, $\Omega(t) = \Omega t$, whereas $\alpha$ and $\omega_{\rm Bohr}$ are $t$-independent constants. On the other hand, the oscillating Schwinger Hamiltonian~\cite{comp,sch}
\[ H_{\rm S}(t)
= \frac{\hbar\omega_{\rm Bohr}(t)}{2}\begin{pmatrix}
\cos\alpha(t) & e^{-i\Omega(t)}\sin\alpha(t)  \\
e^{i\Omega(t)}\sin\alpha(t) & -\cos\alpha(t)   
\end{pmatrix},\] 
and $\Omega$, $\alpha$ and $\omega_{\rm Bohr}$ are all time-dependent variables. We again assume the two energy levels do not cross and there is no $t$-dependent orthogonalization. Here, Fig.~\ref{fig:2} schematically captures the physical system represented by $H_{\rm S}(t)$. In this case, we have $\alpha(t)$ = $\alpha(t)_1$, $\alpha(t)_2$,$\cdots$; $\dot{\alpha}$ = $\dot{\alpha}_1$,$\cdots$; $\Omega(t)$ = $\Omega(t)_1$,$\cdots$; $\dot{\Omega}$ = $\dot{\Omega}_1$,$\cdots$; $\omega_{\rm Bohr}(t)$ = $\omega_{\rm Bohr}(t)_1$,$\cdots$; $\dot{\omega}_{\rm Bohr}(t)$ = $\dot{\omega}_{\rm Bohr}(t)_1$,$\cdots$. This means that one will obtain many $\alpha(t)$, $\dot{\alpha}$, $\Omega(t)$, $\dot{\Omega}$, $\omega_{\rm Bohr}(t)$, $\dot{\omega}_{\rm Bohr}(t)$, and so on due to $t$-dependent fluctuations in the rotating magnetic field (see Fig.~\ref{fig:2}). Now, to guarantee QAA for $H_{\rm S}(t)$, we just need to invoke Eq.~(\ref{eq:x2}) such that $\omega_{\rm Bohr}(t)_x \gg |\dot{\alpha}_x|$, $|\dot{\omega}_{\rm Bohr}(t)_y| \gg |\ddot{\alpha}_y|$, $\cdots$ and $\omega_{\rm Bohr}(t)_u \gg \Omega_u$, $|\dot{\omega}_{\rm Bohr}(t)_v| \gg |\dot{\Omega}_v|$, $\cdots$ where $\{x,y,u,v\} = \mathbb{N}^*$. Using these inequalities, we prove that the two conditions 

\begin {eqnarray}
&&\left|\frac{\dot{\Omega}\sin\alpha - i\dot{\alpha}}{\dot{\Omega}\cos\alpha - \omega_{\rm Bohr} - \frac{\rm d}{{\rm d}\textit{t}}\rm{arg}(\dot{\Omega}\sin\alpha-\textit{i}\dot{\alpha})}\right| = \left|\frac{\Omega'}{\delta'}\right|\ll 1, ~~~{\rm and}~~~\label{eq:xxx1} 
\end {eqnarray}
\begin {eqnarray}
\int_{0}^{\textit{t}}\left|\frac{\rm d}{\rm d\textit{t}'}\frac{\Omega'}{\delta'}\right|\rm d\textit{t}' \ll 1, \label{eq:xxx2}
\end {eqnarray}
derived by Comparat~\cite{comp} are actually the extended versions of Eq.~(\ref{eq:x2}).\\
\textbf{Proof}: For $H_{\rm G}(t)$, Eq.~(\ref{eq:xxx1}) reduces to 
\begin {eqnarray}
&&\left|\frac{\Omega\sin\alpha}{\Omega\cos\alpha - \omega_{\rm Bohr}}\right| \ll 1, \label{eq:xxx3} 
\end {eqnarray}
because $\dot{\alpha} = 0 = (\rm d/{\rm d}\textit{t})\rm{arg}(\Omega')$, $\Omega(t) = \Omega t$ and $\dot{\Omega} = \Omega$. Equation~(\ref{eq:xxx3}) requires $\omega_{\rm Bohr} \gg \Omega$ for QAA to be valid, in accordance with Eq.~(\ref{eq:x2}). We now leave Eq.~(\ref{eq:xxx1}) intact and for QAA to be valid in $H_{\rm S}(t)$, one requires $\omega_{\rm Bohr} \gg |\dot{\alpha}|$, $\omega_{\rm Bohr} \gg \Omega$ and $|\dot{\omega}_{\rm Bohr}| \gg |\dot{\Omega}|$ in which, the first two inequalities satisfy $T_{\rm in} \ll T_{\rm ex}$, while the last one is equivalent to $|\dot{T}_{\rm in}| \ll |\dot{T}_{\rm ex}|$. If $(\rm d/{\rm d}\textit{t})\rm{arg}(\Omega') < 0$ in Eq.~(\ref{eq:xxx1}), then one further requires $|\dot{\omega}_{\rm Bohr}| \gg |(\rm d^2/{\rm d}\textit{t}^2)\rm{arg}(\Omega')|$ or $|\ddot{T}_{\rm in}| \ll |\ddot{T}_{\rm ex}|$, which guarantees $\omega_{\rm Bohr} \gg |(\rm d/{\rm d}\textit{t})\rm{arg}(\Omega')|$ as required from Eq.~(\ref{eq:xxx1}). Having proven that, it is now straightforward to generalize Eq.~(\ref{eq:xxx2}) to obtain
\begin {eqnarray}
\sum_{\textbf{q} = x,y,u,v}\bigg[\bigg(\frac{T_{\rm in}}{T_{\rm ex}}\bigg)_\textbf{q} + \bigg(\frac{|\dot{T}_{\rm in}|}{|\dot{T}_{\rm ex}|}\bigg)_\textbf{q} + \bigg(\frac{|\ddot{T}_{\rm in}|}{|\ddot{T}_{\rm ex}|}\bigg)_\textbf{q} + \cdots\bigg] \ll 1. \label{eq:xxx4} 
\end {eqnarray}
Therefore, Eqs.~(\ref{eq:xxx1}) and~(\ref{eq:xxx2}) are indeed the extended versions of Eq.~(\ref{eq:x2}) $\blacksquare$ 

\subsection*{Example 3: Two-level system with a single energy level crossing (logical exposition)}

For simplicity, we confine ourselves to a two-level system with one energy level crossing occurred somewhere between $t_0$ and $t_1$, say $t_z$. Again, we assume there is no $t$-dependent orthogonalization. In this case, the energy level crossing occurs as a result of some $t$-dependent Yukalov (non-linear)-type interaction, $V(t)$~\cite{yuka}. Hence, one can write the solved Hamiltonian for such a system as
\begin {eqnarray}
&&H(t)\varphi_m(t) = (H_0(t) + V(t))\varphi_m(t) = (h_a(t) + v_a(t))\varphi_m(t) = E_a(t)\varphi_m(t), \nonumber \\&& H(t)\varphi_n(t) = (H_0(t) + V(t))\varphi_n(t) = (h_b(t) + v_b(t))\varphi_n(t) = E_b(t)\varphi_n(t), \label{eq:xxx5} 
\end {eqnarray}
where $H_0(t)$ is the basic Hamiltonian excluding all interactions, whereas $V(t)$ contains all the interaction terms, and we do not require $\langle \varphi_n(t)|\varphi_m(t)\rangle = 0$ or 1. Here, $E_a(t)$, $h_a(t)$, $v_a(t)$, $h_b(t)$, $v_b(t)$ and $E_b(t)$ are all eigenvalues. Degeneracy due to energy level crossing implies $E_a(t_z) = E_b(t_z)$ at a certain point of time ($t_z$), and for other times, they are not degenerate. If they are always degenerate, then $h_a(t) = h_b(t)$, $v_a(t) = v_b(t)$ and $E_a(t) = E_b(t)$, which physically mean $\varphi_n(t) = \varphi_m(t)$: this means that there are two electrons occupying the same energy level for all time $t$. As a consequence, degeneracy due to energy level crossing requires $h_a(t_z) \neq h_b(t_z)$, $v_a(t_z) \neq v_b(t_z)$ and $E_a(t_z) = E_b(t_z)$. Therefore, for degenerate systems, one needs to employ Eq.~(\ref{eq:x2}) such that $1/T_{\rm ex}(t_z) \ll (h_a(t_z) - h_b(t_z))/\hbar$ and $1/T_{\rm ex}(t_z) \ll (v_a(t_z) - v_b(t_z))/\hbar$ where $(v_a(t_z) - v_b(t_z))/\hbar$ and $(h_a(t_z) - h_b(t_z))/\hbar$ equal $1/T_{\rm in}(t_z)$. This is the reason why QAA can still be applied for gapless (due to energy level crossings) systems without strictly requiring $T_{\rm ex} \rightarrow \infty$.  
\subsection*{Example 4: Systems with and without the time-dependent orthogonalization: applied to chemical reactions}

In the previous examples, we have shown how and why QAA is valid for gapped and gapless systems provided that they satisfy Eqs.~(\ref{eq:77.1new1}) and~(\ref{eq:x2}). These conditions are also shown to be necessary and sufficient. In the following example however, we will first provide the logical expositions of how Eq.~(\ref{eq:77.1new}) due to $t$-dependent orthogonalization can render the QAA valid. All we need to do here is to invoke Eq.~(\ref{eq:77.1new}), which now implies  
\begin {eqnarray} 
\frac{\left\langle \varphi_{\rm down}(t_1)\left|\dot{H}(t)\right|\varphi_{\rm up}(t_0)\right\rangle}{E_{\rm up}(t_0) - E_{\rm down}(t_1)} \propto \frac{1}{T_{\rm ex}'(t)} \approx 1. \label{eq:xxx6}
\end {eqnarray}
Equation~(\ref{eq:xxx6}) tells us that the system has started out with spin-up, $\varphi_{\rm up}(t_0)$ at $t = t_0$, and after $t$-dependent orthogonalization, it ends up with spin-down, $\varphi_{\rm down}(t_1)$ at $t = t_1$ where $t_1 > t_0$. For QAA to be valid, we require large external disturbances such that $T_{\rm ex}'(t) \approx 1$ or $T_{\rm ex}'(t) \approx T_{\rm in}'(t)$, which is now necessary for the validity of QAA. Without this large disturbances (or rapidly oscillating Hamiltonian), the spin-up electron (at $t = t_0$) cannot occupy the orthogonalized spin-down state at $t = t_1$. In this case, $c(t_1) \ll 1$ (invalid QAA) as proven in Eq.~(\ref{eq:77.17new}) and explained in the paragraph after Eq.~(\ref{eq:77.17new}). \\
\textbf{Claim}: However, one should note here that Eq.~(\ref{eq:77.1new}) is also applicable for systems without the $t$-dependent orthogonalization. \\
\textbf{Proof}: Recall Eq.~(\ref{eq:77.16}) with $\delta_{t_1t_0} = 1$
\begin {eqnarray}
&&\dot{c}_m(t_1) = -c_m(t_1)\left\langle \varphi_m(t_1)|\dot{\varphi}_m(t_1)\right\rangle - \nonumber \\&& \sum_{t_0 \neq t_1} c_m(t_0)\frac{\left\langle \varphi_m(t_1)\left|\dot{H}\right|\varphi_m(t_0)\right\rangle - \dot{E}_m(t_0)}{E_m(t_0) - E_m(t_1)} e^{i[\theta_m(t_0)-\theta_m(t_1)]}. \label{eq:77.16new}
\end {eqnarray}    
Here, $\delta_{t_1t_0} = 1$ implies that there is no $t$-dependent orthogonalization. First we neglect the second term by requiring $\langle\varphi_m(t_1)|\dot{H}|\varphi_m(t_0)\rangle - \dot{E}_m(t_0)$ is extremely small and/or $E_m(t_0) - E_m(t_1)$ is extremely large, hence  
\begin {eqnarray}
\dot{c}_m(t_1) = -c_m(t_1)\left\langle \varphi_m(t_1)|\dot{\varphi}_m(t_1)\right\rangle. \label{eq:77.16new1}
\end {eqnarray}    
Its solution 
\begin {eqnarray}
c_m(t_1) = c_m(t_0)e^{i\gamma_m(t_1)}, ~~~{\rm where}~~~ \label{eq:77.16new2}
\end {eqnarray}    
\begin {eqnarray}
\gamma_m(t_1) = i\bigg[\int^t_0\left\langle \varphi_m(t')\bigg|\frac{\partial}{\partial t'}\varphi_m(t')\right\rangle dt'\bigg]_{t = t_1}. \nonumber
\end {eqnarray}    
Since the electron occupied the $m^{\rm th}$ eigenstate at $t = t_0$, thus $c_m(t_0) = 1$, and when $t = t_1$, $c_m(t_1) = 1$, and so on. Therefore, we can now recall Eq.~(\ref{eq:77.9}) and write  
\begin {eqnarray}
\Psi_m(t_0) = c_m(t_0)\varphi_m(t_0) e^{i\theta_m(t_0)}, \label{eq:77.9new}
\end {eqnarray}
and using Eq.~(\ref{eq:77.16new2})
\begin {eqnarray}
\Psi_m(t_1) = c_m(t_1)\varphi_m(t_1) e^{i\theta_m(t_1)} = c_m(t_0)\varphi_m(t_1) e^{i\gamma_m(t_1)} e^{i\theta_m(t_1)}. \label{eq:77.9new1}
\end {eqnarray}
Since $\delta_{t_1t_0} = 1$, Eq.~(\ref{eq:77.9new1}) reads
\begin {eqnarray}
\Psi_m(t_1) = c_m(t_0)\varphi_m(t_0) e^{i\gamma_m(t_1)} e^{i\theta_m(t_1)}, \label{eq:77.9new2}
\end {eqnarray}
as it should be, picking up only an additional phase factor ($e^{i\gamma_m(t_1)}$). If the second term on the right-hand side of Eq.~(\ref{eq:77.16new}) is not extremely small, then we strictly require
\begin {eqnarray} 
\frac{\left\langle \varphi_m(t_1)\left|\dot{H}\right|\varphi_m(t_0)\right\rangle - \dot{E}_m(t_0)}{E_m(t_0) - E_m(t_1)} \propto \frac{1}{T_{\rm ex}'(t)} \approx 1, \label{eq:77.9new3}
\end {eqnarray}
in accordance with Eq.~(\ref{eq:xxx6}) $\blacksquare$

We anticipate that the evolution of the eigenfunction in the presence of, or in the absence of $t$-dependent orthogonalization is relevant to understand chemical reactions beyond the standard procedures of calculating activation energies, including the H$_2^+$ molecule ion dissociation~\cite{lefeb} and the water molecule splitting~\cite{water}. For example, chemical reactions (associations or dissociations) strictly require large external disturbances that give rise to significant changes to the eigenfunctions to form new compounds or to dissociate to its constituent chemical components~\cite{god}. In other words, without the significant changes to the eigenfunctions due to large external disturbances, which define the existence of activation energies, the chemical reactions cannot occur. For instance, bringing an atomic hydrogen and H$^{+}$ together chemically to produce H$_2^+$ requires a new wave function, $\varphi(t,\textbf{r})_{\rm new}$ for the ground state of H$_2^+$ molecule. We take the wave function for the atomic H as $\varphi_0(t,\textbf{r})$. Thus, the condition given in Eq.~(\ref{eq:77.1new}) considers the evolution of the wave function, from $\varphi_0(t,\textbf{r})$ to $\varphi(t,\textbf{r})_{\rm new}$. However, the generator ($\hat{\mathcal{G}}$) defined by $\varphi(t,\textbf{r})_{\rm new} = \hat{\mathcal{G}}\varphi_0(t,\textbf{r})$ is not known, and presently $\mathcal{G}$ stands for guess as we often do exactly that to obtain the new wave function by means of the linear combination of atomic orbitals and/or by requiring any arbitrary wave function to give solutions that are convergent. Anyway, we can now invoke Eq.~(\ref{eq:77.1new}), which is also suitable to check the applicability of the QAT for the chemical reaction, H + H$^+$ $\rightarrow$ H$_2^+$. In this case, Eq.~(\ref{eq:77.1new}) can be written as            
\begin {eqnarray}
&&\left|\frac{\left\langle \varphi(t,\textbf{r})_{\rm new}\left|\dot{H}^{\rm chemical}_{\rm reaction}\right|\varphi_0(t,\textbf{r})\right\rangle - \dot{E}_{m}(\rm H)\langle \varphi(\textit{t},\textbf{r})_{\rm new}|\varphi_0(\textit{t},\textbf{r})\rangle}{E_{m}(\rm H) - \textit{E}_{\textit{m}}(\rm H_2^+)}\right| \approx \dot{c}_m(t_k) \approx 1, \nonumber \\&& \label{eq:77.1new2}
\end {eqnarray}
where $E_{m}(\rm H)$ and $E_{m}(\rm H_2^+)$ are the ground state energies of atomic H and H$_2^+$ molecule, respectively. The magnitude of approximately one stated in Eq.~(\ref{eq:77.1new2}) can be achieved during chemical reactions because large external disturbances (by means of thermal and/or potential energies) are required to initiate any chemical reaction. Such a disturbance will first give rise to excitations of the ground state electrons and therefore, QAA is also applicable for such a dynamical process, provided that we know the Hamiltonian ($H(t)^{\rm chemical}_{\rm reaction}$) describing this chemical process explicitly. On the contrary, Eq.~(\ref{eq:77.1new1}) is not applicable for this process, and as we have stated earlier in the introduction, this does not imply that the QAT has been violated just because $\varphi(t,\textbf{r})_{\rm new}$ differs significantly (beyond the phase factor) from $\varphi_0(t,\textbf{r})$.

\section{Further Analysis}

In the discussion and proofs given above, we should not be falsely led into thinking that the condition given in Eq.~(\ref{eq:77.1new}) imposes orthogonality. On the contrary, Eq.~(\ref{eq:77.1new}) allows the final eigenfunction to remain the same, or acquire some phase factors or be mathematically different from the initial eigenfunction, or even be different such that the final eigenfunction is orthogonal to the initial one. Moreover, the proof for Eq.~(\ref{eq:77.1new}) also correctly exposes the intrinsic reason why we need to guess the eigenfunction, for instance, we have used the linear combination of the atomic eigenfunction in the form of Eq.~(\ref{eq:77.9}). In other words, we need to understand what is the mechanism involved, which gives us the license to rewrite $\varphi_0(t,\textbf{r})$ as $\varphi(t,\textbf{r})_{\rm new}$. Well, of course $\varphi(t,\textbf{r})_{\rm new}$ is one of the acceptable solution to the Hamiltonian that gives the lowest ground state energy. But we still lack the fundamental knowledge that enabled us to construct $\varphi(t,\textbf{r})_{\rm new}$ \textit{by hand}. 

For example, we reconsider a H$_2^+$ molecular ion such that the single electron molecular orbital, $\varphi(t,\textbf{r})_{\rm new}$ can be constructed, either from the molecular-orbital (MO), or valence-bond (VB) theory~\cite{god}. In MO theory one uses the linear combination of atomic orbitals (LCAO) approach to construct the molecular orbital, namely, $\varphi(t,\textbf{r})_{\rm new}$ is constructed from the atomic orbitals of the type, $\varphi_0(t,\textbf{r})$. In this case, one immediately starts the construction in the molecular framework, meaning, in the presence of both H and H$^+$. In VB theory, one still uses the atomic orbitals, but they are combined non-linearly~\cite{god}. Therefore, in both cases, the constructions of $\varphi(t,\textbf{r})_{\rm new}$ are carried out in the molecular framework~\cite{god}, even though they are generated from the atomic orbitals in different ways.  

Now, the chemical reaction between H and H$^+$, physico-chemically means that we are bringing two isolated atomic H and H$^+$ ion closer from infinity, or they may come closer on their own due to some attraction between them via the Coulomb interaction at finite distances. Either way, once they are close enough, and with sufficient external heat ($Q$), but not higher (than the H$-$H$^+$ bond energy), the single electron in H can be excited (or polarized) to the highest possible atomic energy level and binds to the H$^+$ ion, to form H$_2^+$. When one reads the above chemical reaction carefully, two intrinsic phenomena pop out that can be written as two statements. 

\textbf{Statement 1}: There is a change in the mathematical property of the wave function (beyond the phase factor) when $\varphi_0(t,\textbf{r})$ for H transforms to $\varphi(t,\textbf{r})_{\rm new}$ for H$_2^+$. 

\textit{Remark 4}: Let us call this the wave function transformation, and this transformation is one form of electronic phase transition arising due to the transition from the atomic (or ionic) to molecular system. The formal proof for the existence of such a quantum phase transition in many-body systems is given elsewhere~\cite{qpt}.

\textbf{Statement 2}: In both MO and VB theories, $\varphi(t,\textbf{r})_{\rm new}$ have been constructed based on the assumption that the chemical reaction has already taken place, since the construction of wave functions were carried out within the molecular framework. 

\textit{Remark 5}: This means that, we are lack of the information needed to find the generator ($\mathcal{G}$) pointed out earlier, and as anticipated, $\mathcal{G}$ still remains as a guess, while to those who are not comfortable with this, we may call it ``an educated guess''. 

In this work, we have actually proven these two statements by proving the existence of the condition stated in Eq.~(\ref{eq:77.1new}). For example, we can observe that Eq.~(\ref{eq:77.1new}) correctly captures the effect of the chemical reaction by requiring an electronic phase transition (due to wave function transformation), in which, the existence of this transition justifies why Remark 2 is necessary. Simply put, even though time is a continuous parameter, we cannot construct a continuous $t$- or $\textbf{r}(t)$-dependent eigenfunction for an electron that starts as an atomic orbital of an isolated atomic H that goes through an electronic phase transition to form H$^+_2$. Here, $\textbf{r}(t)$ denotes the $t$-dependent electron coordinate, which means, we do not require $\varphi(t,\textbf{r})_{\rm new}$ to be both $\textbf{r}$ and $t$ dependents explicitly. 

As a consequence, just because $\varphi(t,\textbf{r})_{\rm new}$ cannot be written as a continuous function of $\textbf{r}(t)$ (starting from isolated H and H$^+$, until the formation of H$_2^+$), one should not assume that the above-stated wave function transformation (or phase transition) never took place, or should be ignored. The very moment we combine the atomic orbitals (linearly or otherwise) within the molecular framework proves the existence of this transformation, which we discovered from Eq.~(\ref{eq:77.1new}). Moreover, Eq.~(\ref{eq:77.1new}) also leads us to understand why the activation energy for the above chemical reaction exists$-$ because it is needed to initiate the wave function transformation (non-observable) or the electronic phase transition (observable). This is consistent with the finite-temperature quantum phase transitions proven in Ref.~\cite{qpt}.  

Finally, before we conclude, let us recall the Born-Oppenheimer (BO) arguments given by Teufel~\cite{ste}. Here, Eq.~(\ref{eq:77.1new}) does not apply for an obvious reason$-$ a given wave function in a particular quantum system, that does not go through a quantum phase transition or $t$-dependent orthogonalization, does not transform beyond the phase factor with time. Therefore, we just need to invoke Eq.~(\ref{eq:77.1new1}) and Eq.~(\ref{eq:x2}) for such systems, which can be done within these well-known theories, MO, VB and the Density Functional Theory (DFT). Briefly, BO approximation is valid for as long as the electrons average kinetic energy equals, or larger than that of the ions, $\frac{1}{2}mv^2_{\rm el} \geq \frac{1}{2}Mv^2_{\rm ion}$, which in turn implies $v_{\rm ion} \ll v_{\rm el}$ and $T_{\rm ex} \gg T_{\rm in}$ if $M \gg m$. Obviously, the above assumption does not violate Eq.~(\ref{eq:x2}), hence QAA is applicable. 

It is also true that repulsive interaction (due to electron-electron interaction) can be invoked to explain the existence of finite energy gap~\cite{ste,ber}, however, energy levels do cross due to ``lucky coincidences'' in many-body systems, as shown in Example 3. Even in the presence of such coincidences, we can still check the applicability of QAA via Eq.~(\ref{eq:x2}). Now, in the worst case scenario, we will have (refer to Example 3)
\begin {eqnarray}
h_a(t) = h_b(t),~~ v_a(t) = v_b(t) ~~{\rm and}~~ E_a(t) = E_b(t), \label{eq:new}
\end {eqnarray}
then QAA seems to be not applicable. But if we look closely, Eq.~(\ref{eq:new}) physically imply $\varphi_n = \varphi_m$. Hence, one can indeed invoke Eq.~(\ref{eq:x2}) through the BO approximation and justify the applicability of QAA because the transition between $\varphi_n$ and $\varphi_m$ does not violate QAA. Of course, mathematically one can construct the eigenfunctions such that $\varphi_n \neq \varphi_m$ and still satisfy Eq.~(\ref{eq:new}). But this is like saying, we are attempting to invalidate QAA, solely for the purpose of violating QAA without any physical justification.

\section{Conclusions}

In conclusion, we have developed a new version of the quantum adiabatic theorem that takes the time-dependent orthogonalization into account, and even in the absence of this orthogonalization, we found that these (with and without the $t$-dependent orthogonalization) effects are very relevant to study the dynamics of chemical reactions. We also generalized and incorporated the relationship between the internal and external timescales as an additional, but separate condition into the quantum adiabatic theorem. The time-dependent orthogonalization and chemical reactions are proven to be captured by Eq.~(\ref{eq:77.1new}) self-consistently. In addition, Eq.~(\ref{eq:77.1new}) also led us to discover the activation energy in a given chemical reaction is needed for the wave function transformation, which can be observed via the electronic phase transition. While the validity of other conditions are explicitly exposed to be sufficient in the spin-1/2 two-level system. Therefore, the new theorem is shown to be valid for both gapped and gapless systems with appropriate examples. 

For gapped two-level system, Eqs.~(\ref{eq:77.1new1}) and~(\ref{eq:x2}) are found to be sufficient and necessary, in accordance with Comparat~\cite{comp}. Whereas, for gapless systems with degeneracies (due to energy level crossings) and quantized energy levels, Eqs.~(\ref{eq:77.1new1}) and~(\ref{eq:x2}) are again sufficient and necessary, in agreement with Avron and Elgart~\cite{avron}. However, for gapless systems, we need not keep track of the eigenfunctions to show they are sufficient, for as long as Eq.~(\ref{eq:x2}) is valid. Interestingly however, we do not always require $T_{\rm ex} \rightarrow \infty$ in the presence of non-linear Yukalov-type~\cite{yuka} interactions. The importance of Eq.~(\ref{eq:x2}) cannot be overstated as the recent counter-examples~\cite{comp,ms} and counter-proofs~\cite{comp,sab,ms} can be traced back to the violation of Eq.~(\ref{eq:x2})~\cite{unp}.   

\section*{Acknowledgments}

This work was supported by Sebastiammal Innasimuthu, Arulsamy Innasimuthu and Arokia Das Anthony, as well as the Ad-Futura visiting researcher fellowship (Slovenia) between Feb/2010 and Feb/2011. I also would like to thank the School of Physics, University of Sydney for the USIRS award (2007$-$2009) where part of this work was carried out, and the University of Sydney Library for providing e-access to some of the references. I am grateful to several anonymous readers and referees for correcting my mistakes and for introducing me to the physics of exceptional points. Special thanks to the Progress of Theoretical Physics (Kyoto) editorial committee for waiving the publication fee.

\end{document}